\documentclass[preprint]{aastex}

\newcommand{\boxpap}{BOX}

\slugcomment{to appear in ApJS v145, March 2003}

\shorttitle{Tree--Particle--Mesh}
\shortauthors{Bode \& Ostriker}

\received{2002 July 22}
\begin{document}

\title{ Tree--Particle--Mesh: an adaptive, efficient, and parallel 
code for collisionless cosmological simulation
}

\author{Paul Bode and Jeremiah P. Ostriker}
\affil{Princeton University Observatory, Princeton, NJ 08544-1001}
\email{bode@astro.princeton.edu, jpo@astro.princeton.edu}

\begin{abstract}
An improved implementation of an N-body code for simulating
collisionless cosmological dynamics is presented.  TPM
(Tree--Particle--Mesh) combines the PM method on large scales with
a tree code to handle particle-particle interactions at small
separations. After the global PM forces are calculated, spatially 
distinct regions above a given density contrast are located; the 
tree code calculates the gravitational interactions inside these 
denser objects at higher spatial and temporal resolution.  
The new implementation includes individual particle 
time steps within trees,  an improved treatment of tidal forces
on trees, new criteria for higher force resolution and choice of 
time step, and parallel treatment of large trees.  TPM is compared
to P$^3$M and a tree code (GADGET) and is found to give equivalent 
results in significantly less time.  The implementation is highly 
portable (requiring
a Fortran compiler and MPI) and efficient on parallel machines.
The source code can be found at 
\url{http://astro.princeton.edu/$\sim$bode/TPM}.
\end{abstract}


\section{Introduction} \label{sec:intro}

Numerical simulation of the nonlinear evolution of collisionless 
dark matter, usually by following a set of point masses as they 
move under their mutual gravitational influence, has, over the last 
few decades, played an important role in increasing our understanding 
of cosmological structure formation.
Rather than attempt to make a summary of the extensive
literature available on this subject here, we
direct the reader to the recent reviews by \citet{Bert98}
and \citet{Klypin00}.

A search of the internet will reveal a diverse set of N-body codes 
now available to be downloaded
by those interested in carrying out cosmological simulations. 
One of the first to be developed is the direct code of \citet{Aars99}.
The operation count for implementing any such particle--particle code 
scales as $N^2$, where $N$ is the number of particles.
One can reach larger $N$ and thus higher mass resolution (but spatial 
resolution limited by the cell size on a grid) 
in the same amount of CPU time using 
the PM (Particle--Mesh) method, for example the code by \citet{KlypHol97}.
The operation count here scales as $N$log$N$, with a small prefactor;
PM is implemented on regular grids, where 
efficient fast Fourier transform (FFT) algorithms are available.
To attain higher, subgrid resolution one can add to a PM code
particle--particle interactions and refined subgrids, such as in the
Hydra code \citep{CouchTP95}.  A gridless approach is used in 
tree codes, for example the ZENO package of \citet{Barnes98}.
The operation count in the Barnes--Hut tree algorithm \citep{BarnHut86}
also scales as $N$log$N$ but the prefactor is much larger---
by a factor of 10--50 \citep{Hern87}--- than for a PM code.
The tree code of \citet{Hern87} has been made parallel \citep{BAD01}.
A tree code with individual particle time steps called GADGET has
recently been released in both serial and distributed memory parallel 
versions \citep{SpringelYW01}.   One promising new development is 
the use of multigrid methods involving mesh refinements of arbitrary 
shape \citep{KKK97}, such as in MLAPM \citep{KnebeGB01}.
Another cell-based approach has recently been developed which 
shows improved performance over standard tree codes 
\citep{Dehnen00,Dehnen02} and is included as part of the 
NEMO package \citep{Teuben95}.

For a given choice of algorithm, a related problem is making it run 
efficiently on parallel computers \citep{YaMoYo99, RiDoLa00, VitCar00, 
LiaCar01, SpringelYW01, YaYo01, DoHeMe02, JingSuto02, MioCD02, Stadel02}.
Gravity is long-range, so to compute the force on a given particle, one
needs some information from every other particle, and communication 
is thus required if particles are distributed among many processors.
\citet{Xu95} presented a new algorithm, called Tree--Particle--Mesh,
based on using domain decomposition in a manner
designed to overcome this difficulty.  TPM uses the efficient PM
method for long-range forces and a tree code for sub-grid resolution.
Isolated, overdense regions are each treated with a separate tree,
thus ensuring coarse--grained parallelism. 

\citet{BOX00}, hereafter \boxpap , made several improvements in
the code of \citet{Xu95}.  In this paper we present further refinements
which improve the accuracy and efficiency of the algorithm.
These include the following: allowing individual particle time steps within
trees,  an improved treatment of tidal forces, new criteria
for higher force resolution and for the choice of time step, and parallel
treatment of large trees.  In \S\ref{sec:overview} we present an
overview of latest version of TPM, including time stepping;
\S\ref{sec:implementation} discusses implementation of domain
decomposition and the particle time step criterion, with an
overview of how the code works in practice.  The accuracy of
the code is explored in \S\ref{sec:perfandacc} by comparing results
with two other algorithms: P$^3$M and the GADGET tree code.
Concluding remarks and details of the public release
of the TPM code are given in \S\ref{sec:discussion}.

\citet{Calder02} discuss the distinction between verifying a simulation
(that is, being sure that the equations are being solved
correctly) and validating it (having confidence that the equations
and their solution actually resemble a real world problem).
Obviously, in this paper only the former is done.  Various aspects
of the validity of cosmological simulations are addressed in \citet{Klypin00},
\citet{KnebeKGK00}, \citet{vanK00}, \citet{HaYoSu02}, \citet{Knebe02}, 
\citet{Powetal02}, \citet{BaeJoLa02}, and references therein.
When using an N-body code, the simulator will need to be
cognizant of these issues.

\section{Overview of TPM} \label{sec:overview}

TPM begins with a standard PM code (in fact, if there are no particles
selected to be in trees, then TPM defaults to a PM code);
this portion is similar to the PM code described by \citet{Efst85}.
The density on a regular grid is found by interpolating
particle positions using the cloud--in--cell (CIC) method,
and Poisson's equation is solved on this grid using 
Fast Fourier Transforms.  As noted in the introduction, 
the FFT technique is highly efficient and scales 
as $N$log$N$ \citep{HockEast81}.
In order to obtain subgrid resolution a tree code is used.
One possible manner of doing this is to compute shorter range
forces for every particle with a tree code \citep{Bagla99},  
in a manner analogous to the P$^3$M algorithm.
Instead, TPM uses domain decomposition,
taking advantage of the fact that potentials due to matter
outside and inside a given domain can be added linearly,
and it uses different solvers to find these two potentials.
This section begins with a very broad description of the TPM
algorithm, and then considers various aspects in more detail.

A basic outline of TPM can be summarized as follows:
\begin{enumerate}
\item Identify tree regions and the particles in each such region;
particles outside of tree regions are evolved with PM only.
\item Push PM particles to midstep 
\item Find the tidal potential in each tree region--- the potential
due to all mass outside of that region.
\item Integrate each tree region forward to the middle of the PM step.
\item Find the PM potential and update the acceleration for PM particles.
\item Integrate each tree region forward to the end of the PM step.
\item Update PM particle velocities and positions to the end of
      the time step.
\end{enumerate}
The motivation behind this algorithm is that the motions of 
particles within a given tree region can be integrated 
independently using high temporal and spatial resolution:
knowledge of the rest of the simulation is not required, 
except for the time--averaged, externally produced tidal forces.

The stepping in time of positions ${\bf x}$ and
velocities ${\bf v}$ is accomplished with a standard second order 
leapfrog integration in comoving coordinates, the cosmological model
determining the scale factor $a(t)$:
\begin{mathletters} \label{eqn:leapfrog}
\begin{eqnarray}
 {\bf x}(t+\onehalf \Delta t) & = & 
	{\bf x}(t) + \onehalf {\bf v}(t)\Delta t  , \\
 {\bf g}(t+\onehalf \Delta t) & = &  -\nabla \Phi({\bf x})  , \\
 {\bf v}(t+\Delta t) & = & 
    \frac{1-H\Delta t}
         {1+H\Delta t}{\bf v}(t)
  + \frac{a^{-3}}
         {1+H\Delta t}{\bf g}(t+\onehalf \Delta t )\Delta t  , \\
 {\bf x}(t+\Delta t) & = & {\bf x}(t+\onehalf \Delta t) + 
     \onehalf {\bf v}(t+\Delta t)\Delta t  .
\end{eqnarray}
\end{mathletters}
Here $a$, $H\equiv \dot{a}/a$, and the gravitational potential $\Phi$
are determined at time $t+\onehalf \Delta t$.

\begin{figure}
\plotone{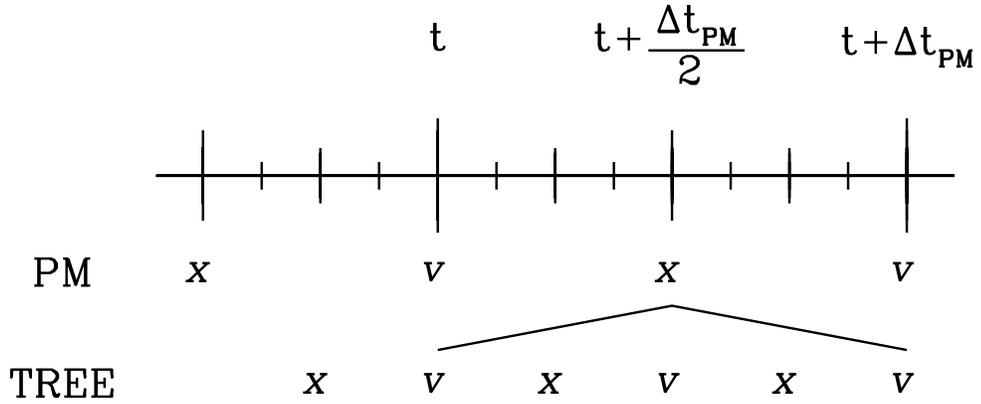}
\caption{Diagram of time stepping: 
velocities are updated at the times marked $v$ and
accelerations at times marked $x$.
\label{fig:tstep}}
\end{figure}

One advantage of the TPM approach is that it allows the use of
multiple time steps.  Tree particles are required to take at
least two steps per PM step, but each particle has an individual
time step so that finer time resolution can be used if required.
Particles are arranged in an hierarchy of time bins differing
by a factor of two, in the manner of \citet{HernKatz89}:
\begin{equation} \label{eqn:timebins}
    \Delta t_{tree} = \frac{\Delta t_{PM}}{2^s} ,
\end{equation}
where $\Delta t_{PM}$ is the PM time step and the integer
$s\geq 1$.  
In a sense TPM operates along the same
lines as a tree code, except that the particles in the
$s=0$ time step bin are handled differently from the rest.
A diagrammatic representation of the time stepping is shown
in Fig$.$\,\ref{fig:tstep} for the case when all tree particles
are in the longest time step bin, $s=1$.
Beginning at time $t$, the PM particle positions are moved forward
to midstep (eq.~[\ref{eqn:leapfrog}a] with $\Delta t =\Delta t_{PM}$).
The tidal potential is then found, and the tree particles are evolved
forward one full step with $\Delta t =\Delta t_{PM}/2$.  The PM
potential is updated (eq.~[\ref{eqn:leapfrog}b]), and for a second time 
the tree particles are evolved for one full step, to time 
$t+\Delta t_{PM}$.  Finally, the PM particle velocities and positions
are updated to the end of the step (eq.~[\ref{eqn:leapfrog}c-d]).

To repeat the TPM algorithm in more detail:

\paragraph{1.} Identify tree regions and the particles in each region.
This is done by identifying PM cells above a given density threshold,
described in \S\ref{sec:domdec}.  Adjoining cells are then
grouped into isolated tree regions, as described in \boxpap .

\paragraph{2.} Push PM particles to midstep (eq.~[\ref{eqn:leapfrog}a]).

\paragraph{3.} Find the tidal potential $\Phi_{ext}$ for each tree region.
First the total potential is
computed on the grid in the standard PM manner.  Then
for each of the tree regions, a small portion of the grid 
containing the PM potential is saved, and the contribution 
to this potential made by the tree itself is subtracted out,
leaving the tidal potential due to all external mass.
See \boxpap\ for details.

Since $\Phi_{ext}$ can only be calculated once per
PM step, it will be calculated at midstep.  PM particles are thus
already in the proper location for this, but tree particles, with 
positions at the beginning of the PM time step, are not.
However, since tree regions are spatially separated and the
halo density profiles are evolving on slower timescales than
any individual particle's orbital period, an approximate position
is sufficiently accurate.  (Note that since a given tree's
own contribution to $\Phi_{ext}$ is exactly subtracted back out,
only the effect on other trees needs to be considered).  Thus, an
approximation of the tree particles' positions at the middle
of the PM step is taken to be the position one full particle time 
step ahead; since tree particle time steps are half the
PM time step or less, this is no more than the PM midstep.
With these advanced positions, the potential on the grid is found
in the standard PM manner.

For each tree the PM potential in a cubical subvolume is saved.
This volume is
slightly larger than the  active cell region (in order to allow for
finding the gradient of the potential through finite differencing)
plus one extra cell on a side in case particles migrate out of
the active cell region during the time integration.  Since the 
PM time step is limited by a Courant condition (see \S\ref{sec:timestep})
such that PM particles cannot move more than a fraction of a cell
per step, one extra cell is sufficient--- particles near the edge
of a tree region are at only a slightly higher overdensity
than the densest PM regions, and hence have similar velocities.
At the beginning of a tree step the portion of the PM potential
due to the tree particles themselves is subtracted out, leaving
the tidal potential.  With $\Phi_{ext}$ saved in this manner,
every time a particle's acceleration is updated, the tidal force is 
calculated from the grid in the same manner that PM forces are found. 
This is an improvement over the method used in \boxpap .

\paragraph{4.} Integrate each tree region forward to the middle of the PM step.
This is done with a tree code, adding in the tidal forces. 
The tree code we use was written by 
Lars Hernquist \citep{Hern87, HernKatz89, Hern90}.
Any other potential solver could be used, but a tree code is
well suited for this type of problem,  in that it
can efficiently handle a wide variety of particle distributions,
can include individual particle time steps,
and scales as $N$log$N$.

Note that for each tree this portion of the code is self--contained;
that is, once the particle data and potential mesh for a given
tree have been collected together no further information is required
to evolve that tree forward.  This makes TPM well suited for parallel
processing on distributed memory systems.  Once the tree data is
received by a given processor this step--- which is the 
most computationally expensive part of the algorithm--- can be performed
without any further communication.  Given this coarse--grained
parallelism, one can use widely distributed and
heterogeneously configured processors, with more capable processors
(or groups of processors) reserved for the largest trees.

\paragraph{5.} Find the density and potential on the PM grid again, 
and update the acceleration for PM particles (eq.~[\ref{eqn:leapfrog}b]).
All tree particles are at the PM midstep, the proper location
for computing the force on PM particles.

\paragraph{6.} Integrate each tree region forward to the end of the PM step.

\paragraph{7.} Update PM particle velocities and positions to the end of
the time step (eqs.~[\ref{eqn:leapfrog}c-d]).

\section{Other Aspects of the Implementation} \label{sec:implementation}

\subsection{Domain Decomposition} \label{sec:domdec}

TPM exploits the fact that gravitational instability in an
expanding universe creates spatially isolated high density peaks;
for typical CDM models these peaks will contain a significant 
fraction of the mass ($\sim 1/4$) but occupy a tiny fraction 
of the volume ($\sim 10^{-3}$; 
see for example Fig$.$\,2 of \citet{MoWhite02}).
As described in detail in \boxpap , this is accomplished by
tagging active mesh cells based on the cell density (which has
already been calculated in order to perform the PM step).
Adjacent cells are linked together, dividing the cells into 
groups in a friends--of--friends manner.  Thus isolated 
high--density regions of space are identified, separated by at 
least one cell width.

In \boxpap , the criterion used to decide if a cell $i$ was
active was if its density $\rho_i$ exceeded a global threshold density
\begin{equation} \label{eqn:oldrhocrit}
    \rho_i \geq \rho_{\rm thr} = A\bar{\rho} + B\sigma ,
\end{equation}
where $\bar{\rho}$ is the mean density and $\sigma$ is the
dispersion of cell densities.
While temporally adaptive (in the sense that the threshold
is lower at early times when peaks are rare),
this criterion has two main drawbacks.  First, small
peaks in isolated regions or voids will not be picked up, 
especially when $\sigma$ is large, because the total mass in
such a halo would not put a PM cell above $\rho_{\rm thr}$.
Secondly, in the case of a region with peak density near
$\rho_{\rm thr}$, whether or not a given PM cell is above
$\rho_{\rm thr}$ can depend on the offset of the grid with
respect to the particles--- the halo mass may be divided
into two or more cells.

We have developed an improved criterion for selecting tree 
regions involving the contrast of a cell with its surroundings 
(rather than a single global threshold value) found by comparing 
the density smoothed on two scales.  For each cell three densities
are computed. First $\rho_s$, found by boxcar smoothing over a
length of 5 cells along each dimension. (Another type of 
smoothing, {\it e.g.} with a Gaussian, would be possible,
but the boxcar can be done with minimal interprocessor 
communication).  Second, $\rho_c$, the mean density in the
eight--cell cube of which the cell under consideration is the 
lower octant (when using CIC, a particle in this cell would
also assign mass to the other cells in this cube). Third, 
$\rho_{\rm ext}$, a measure of the density surrounding this
cube, defined as
\begin{equation} \label{eqn:rhoext}
    \rho_{\rm ext} = \frac{5^3\rho_s - 2^3\rho_c}
			  { 5^3 - 2^3 }.
\end{equation}
All eight cells used to find $\rho_c$ are marked as active
if
\begin{equation} \label{eqn:rhocrit}
    \rho_c \geq \rho_{\rm thr} = \left( A\bar{\rho} + B\sigma \right)
    \frac{\rho_{\rm ext}}{ \bar{\rho} + \sigma + \rho_{\rm ext} }.
\end{equation}

The reasons for such a choice of form can be seen by examining
the behavior of eq.~[\ref{eqn:rhocrit}] in various limits. At
early times in a cosmological simulation 
$\rho \approx \bar{\rho}$ and $\sigma\ll \bar{\rho}$, so 
$\rho_{\rm thr} \approx \onehalf A\rho_{\rm ext} \approx \onehalf A\bar{\rho}$.
Thus with a choice of $A=2$ those regions which are only slightly 
overdense will be selected.  At late times 
$\sigma\gg \bar{\rho}$.  In void regions 
$\rho_{\rm ext}\ll \sigma$, which means 
$\rho_{\rm thr} \approx B\rho_{\rm ext}$;
thus an isolated cell will be chosen if it is a fixed multiple
above its surroundings.  If on the other hand $\rho_{\rm ext}\gg \sigma$,
then $\rho_{\rm thr} \approx B\sigma$, meaning that at each step there is
a given density above which all cells will be chosen; this limit 
will increase with time as $\sigma$ increases. 
Keep in mind that the distribution of cell densities will be 
approximately lognormal
(\citet{KayoTaSu01} and references therein) when interpreting the 
value of $\sigma$.

After cell selection, the domain decomposition proceeds in the manner
described in \boxpap .  Cells are linked by friends--of--friends,
yielding regions of space separated by at least one cell length.
Any particle which contributes some mass to an active cell when
finding the PM density is assigned to the corresponding tree.

Because the volume of all tree regions is less than one percent
of the total volume, the amount of memory allocated to hold the
tidal potential data is generally not significant compared to
that already needed for the PM mesh.  However,
one complication of this scheme is that at early times, when
$\sigma$ is small, the resulting trees can be very long filaments.
While not containing a large amount of matter, being not very
overdense and nearly one--dimensional, the size of a cubical
volume enclosing such a tree will sometimes be a significant fraction of
the entire simulation volume.  This can cause difficulties because
of the amount of computer memory required to 
compute and save the tidal potential in such subvolumes.
The value of $B$ in eq.~[\ref{eqn:rhocrit}] is thus allowed to 
increase at early times when the spatial extent of trees tends
to become larger, and decrease at later times when trees are 
more compact.  We have settled on the following method, which
has worked well in a variety of simulations.
Space is allocated for a maximum subvolume length
one quarter of the PM mesh size.  The size of the largest sub-box
used is checked at the end of each PM step.  At early times 
(when $\sigma<1.6$) $B$ is increased by 0.5\%
if this box has a length more than a third of the allocated value.
Once $\sigma$
exceeds 1.6 in a typical simulation, the spatial extent of trees has
stopped growing, so in this case $B$ is decreased by 0.25\% if the
largest sub-box has a length less than half the maximum allocated.
A minimum value is set for $B$ to prevent too many tree
particles being chosen;  a choice of $B=10$ will place roughly half
of the particles in trees at $z=0$.  An example of this in practice
is discussed in \S\ref{sec:overv}.

The number of PM grid cells can be more or less than the number
of particles, though of course the finer the grid, the greater
the accuracy of the PM and tidal forces.  Generally, the number
of cells should at least equal the number of particles; we have
found that eight grid cells per particle works well.

\subsection{Time Step Criterion} \label{sec:timestep}

A variety of methods for determining the particle time step
have been proposed.  This diversity of criteria reflects
the fact that the appropriate time step can depend on a
number of quite different considerations.  These include
the local density or the local dynamical time, nearby
substructure, the nearest
neighbor, and the softening length \citep{KnebeKGK00, Powetal02}.
In addition, for any single
criterion, one can imagine a special circumstance where it 
breaks down.
This suggests the possibility of
basing $\Delta t$ on more than one criterion.
Thus in the TPM code
we have decided to combine a number of possible criteria, drawn
from variables which do not require any significant computation
to find, in the following manner: 
\begin{mathletters} \label{eqn:deltat}
\begin{eqnarray}
 \Delta t & = & 
   \eta \cdot {\rm MAX}\left( \Delta t_1,\: \Delta t_2 \right) {\rm , where} \\
 \Delta t_1 & = & 
   {\rm MIN}\left( \epsilon /v,\: \sqrt{a^3\epsilon /g} \right) {\rm , and} \\
 \Delta t_2 & = & {\rm MIN}\left( v/g,\: g/\dot{g},\: 
			      \frac{\epsilon^2 + r^2}{rv} \right) .
\end{eqnarray}
\end{mathletters}
Here $\eta$ is an adjustable dimensionless parameter,
$\epsilon$ is the spline kernel softening length, and $r$ is the distance
to the nearest neighboring particle. The latter is found as
a byproduct of the force calculation; if at the beginning of a
tree step $r$ has not been calculated, one can 
as a proxy find the minimum
$r$ that is consistent with the particle's current $\Delta t$.
The time derivative of $g$ is approximated by 
$\dot{g}=|g(t)-g(t-\Delta t_s)|/\Delta t_s$. 
This criterion for $\Delta t$ is quite conservative,
and is similar in principle to that adopted by \citet{Aars99}.

\begin{figure}
\plotone{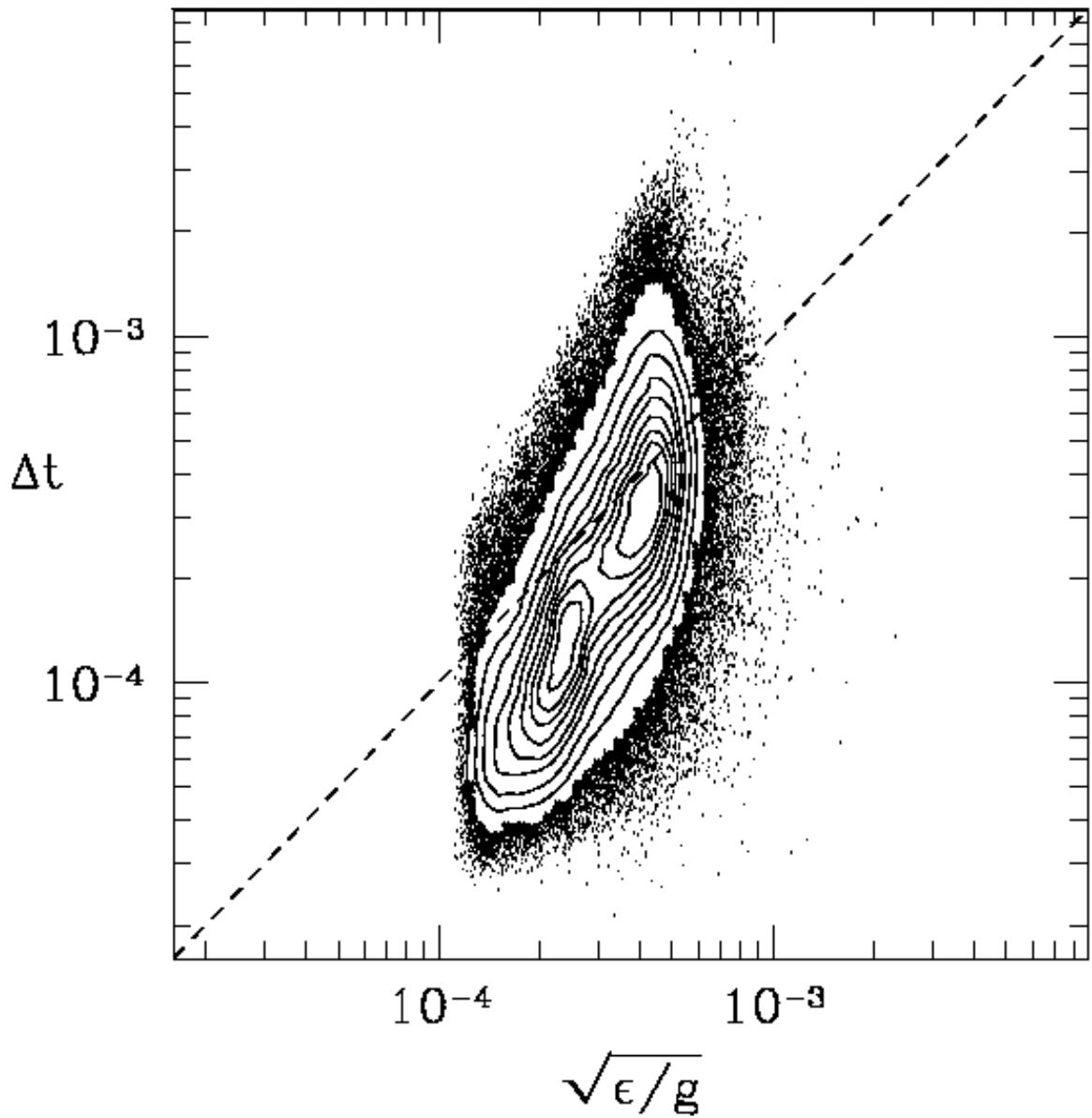}
\caption{Comparison of the time step $\Delta t$ returned by 
eq.~[\ref{eqn:deltat}] with that determined by 
$\eta \sqrt{\epsilon /g}$ for particles in the
largest tree of a cosmological simulation at $z=0$.  Each contour level
contains one tenth of the particles, and the remaining tenth are plotted
as points; the dashed line shows when the two are equal.  Code units
(in which $G$, the total mass, and the box length are all unity) are used,
with $\eta=1$.
\label{fig:dtcomp}}
\end{figure}

A comparison of the time step returned by eq.~[\ref{eqn:deltat}]
and the often used criterion $\eta \sqrt{\epsilon /g}$ is given in
Fig$.$\,\ref{fig:dtcomp}, which shows how these two criteria
compare at $z=0$ for particles in the largest tree of the 
simulation discussed in \S\ref{sec:overv} and
\S\ref{sec:codecomp}.  This tree is made
up of over $1.4\times 10^5$ particles, and contains two large
halos plus a number of smaller satellites and infalling matter.
Each contour level in Fig$.$\,\ref{fig:dtcomp} encloses a tenth of
the particles, and the remaining tenth are plotted as points.
It can be seen that eq.~[\ref{eqn:deltat}] tends to yield a 
smaller timestep than $\eta \sqrt{\epsilon /g}$; this is the case
for 74\% of the particles.  Generally the differences are not
large--- less than a factor of two for 60\% of the particles.
Another trend seen in the figure is that as $\eta \sqrt{\epsilon /g}$
becomes smaller it is more likely for eq.~[\ref{eqn:deltat}] to
give an even smaller value, and conversely eq.~[\ref{eqn:deltat}] 
tends to give a longer time step than $\eta \sqrt{\epsilon /g}$ as the
latter becomes larger.  In most cases the value of $\Delta t$
is set by the factor 
$(\epsilon ^2 + r^2)/(rv) = (r/v)(1+\epsilon ^2/r^2)$.
In other words, the distance to the nearest neighbor is not allowed
to change by a large factor, unless this distance is small
compared to the softening length.  Even in cases when this
factor is larger than $\sqrt{\epsilon /g}$, it is still smaller
than $v/g$ and $g/\dot{g}$;  since the equations being integrated
have the form $\Delta x = v\Delta t$ and $\Delta v = g\Delta t$ 
this limits integration errors.

The PM time step is limited by a Courant condition,
such that no PM particle moves more than one quarter of a cell
size per step.  Also, $a$ is not allowed to change by more
than 1\% per step.  Initially this latter criterion is the most
restrictive, but over time it allows a longer $\Delta t_{PM}$ and 
becomes unimportant.  The $\Delta t_{PM}$ allowed by the Courant
condition also tends to increase over time, although more slowly.
This is because particles falling into dense regions are placed
into trees, and the velocities of the remaining particles are
redshifted (eq.~[\ref{eqn:leapfrog}c]).  The largest time step
for any particle, $\Delta t_{PM}$, is kept
constant until it can safely be increased by a factor of 2, 
and it is then doubled.  The time step for tree particles is
unchanged, which means they take twice as many steps per PM step
as before;  during tree evolution particles will move into a lower 
time step bin if allowed by eq.~[\ref{eqn:deltat}].
Whenever any particle changes time step, its position is updated
as described in \citet{HernKatz89} to preserve second order
accuracy.

\subsection{A typical run} \label{sec:overv}

Various aspects of a typical TPM run can be demonstrated
with a standard cosmological simulation.
This test case contains $N$=128$^3$
particles in a box $L$=40$h^{-1}$Mpc on a side.  The number of
grid points for the PM and domain decomposition portions of
the code is eight times the number of particles, or $256^3$.
The initial conditions were generated using the publicly 
available codes GRAFIC1 and LINGERS\footnote{These codes are 
available at \url{http://arcturus.mit.edu/grafic/}}
\citep{Bert01, MaBert95}.  
A spatially flat LCDM model was
chosen, with cosmological parameters 
close to the concordance model of \citet{OstStein95}:
$\Omega_m=0.3, \Omega_\Lambda=0.7, h=0.70, \Omega_b h^2=0.20,
n=1$, and $\sigma_8=0.95$.  The particle mass is thus 
$2.54\times 10^9h^{-1}M_\odot$.  
Within the tree portion of the code, the opening parameter 
$\theta$ in the standard Barnes--Hut algorithm \citep{BarnHut86} is set to
$\theta = 0.577 \approx 1/\sqrt{3}$, and the time
step parameter $\eta$=0.3 (see eq.~[\ref{eqn:deltat}]).
The cubic spline softening
length was chosen to be 3.2$h^{-1}$kpc so that the
spatial dynamic range $L/\epsilon \sim 10^4$.
With such a small softening length,
halos with fewer than 150 particles are likely to 
undergo some 2-body relaxation in their cores over a Hubble time
(assuming the particle distribution follows an NFW profile with
concentration $c$=12).

\begin{figure}
\plotone{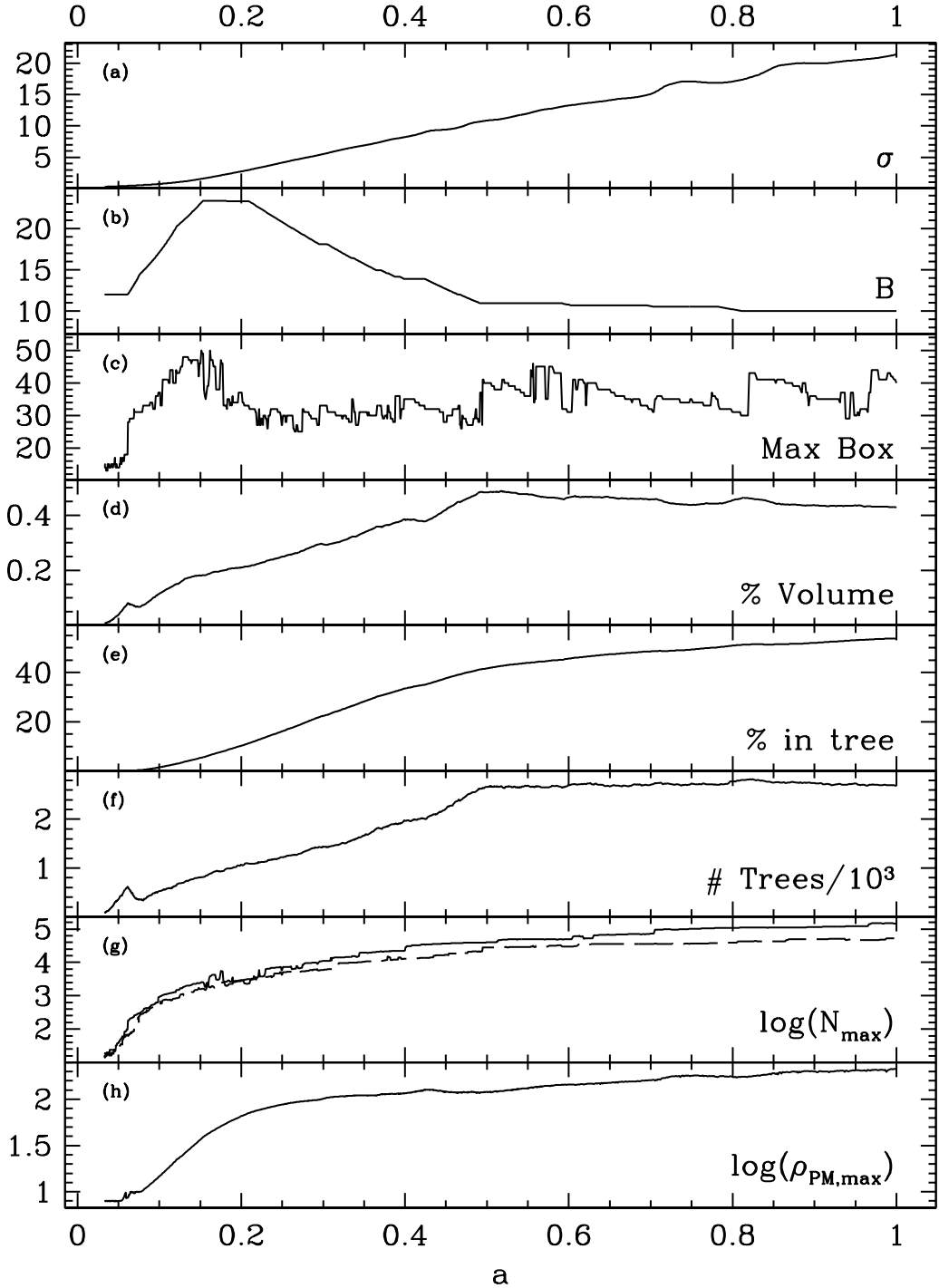}
\caption{Overview of typical run: $(a)$ dispersion of PM cell densities
(where the mean=1),
$(b)$ value of $B$ used in eq.~[\ref{eqn:rhocrit}], 
$(c)$ maximum tree subvolume size,
$(d)$ \% of total volume in trees, $(e)$ \% of all particles in trees,
$(f)$ number of trees (in thousands), 
$(g)$ log of number of particles in the
first and third most massive trees, $(h)$ log of densest PM-only cell. 
\label{fig:overv}}
\end{figure}

Fig$.$\,\ref{fig:overv} displays the evolution of several 
important quantities during this run
as a function of expansion parameter $a$.
The top panel shows the dispersion of PM cell densities $\sigma$ 
(where the mean cell density is unity), and the second panel
shows the value of $B$; both of these factor in
eq.~[\ref{eqn:rhocrit}], which in turn plays a major part in
determining the tree distribution.  Various aspects of this
distribution are shown in the remaining curves.
The third panel shows the size of the largest cubical subvolume
required (in units of PM grid cells).  Initially small, this
grows extremely rapidly as $\sigma$ rises from its initial
value of 0.3 to $\sim 1$; in response $B$ is increased.
When this size is at its greatest (at $\sigma\approx 1.6$), 
the percentages of total volume and mass in trees (shown in
the next two panels) are still quite small.  The trees at
this time tend to follow caustics--- they are only slightly
overdense and not very massive, but because of their filamentary
nature they can have a large spatial extent.  These caustics
then fragment and collapse, so--- even while the total mass in trees
and the number of trees (shown in the third panel from the bottom)
increase--- the maximum subvolume size decreases.
The changes in $B$ affect mainly the maximum subvolume size;  
the number of trees and volume contained
in tree regions both increase at a steady rate up to $z$=1.
The number of tree particles increases monotonically throughout
the simulation.

After $z=1$ (by far the bulk of the computational time),
the characteristics of the simulation change much 
more slowly.  Roughly half the particles are in $\sim$2700 trees
ranging from a few to $10^5$ particles but occupying only 0.4\% of
the simulation volume.  The penultimate panel shows the number of
particles in the first and third largest trees;  by $z=1$ the
growth in mass of these objects has slowed to a fairly constant,
low rate.  Trees are distributed in mass roughly as a power law,
with 67\% of the trees having fewer than 100 particles and
95\% less than 1000.  Most tree particles take two or
four steps per PM step, but some are taking up to 64; there
were 1075 PM steps total in this run.

As $\sigma$ increases, so does the limiting
density above which all cells are treated at full resolution.
It is important to keep track of this limit when interpreting
the results of a TPM run.
The bottom panel shows the highest value of the density among
those cells evolved only with PM.  By $z=1$ this density is 120
times the mean, corresponding to 15 particles inside a cell.
By the end of the computation, the densest PM--only cell
contains 26 particles;  so trying to make statements about
objects smaller than this will be complicated by the varying
spatial resolution of TPM.  Note that most objects with $N\la 25$
will in fact be followed at full resolution--- substructures 
inside larger objects because they are in regions of higher
density, and small isolated halos because they present a
density contrast to their surroundings.
To be cautious we will limit our
analysis to objects 50\% larger than this limit, {\it i.e.} 40 particles
or $10^{11}h^{-1}M_\odot$.

\subsection{Load balancing} \label{sec:loadbal}

As seen in the previous section, during
later epochs--- which take up most of the computational time
needed for a run--- the mass distribution of trees is generally well 
fit by a power law ranging from a few particles up to of
order 0.01$N$.  The actual mass of the largest tree will depend
on the ratio of the largest non-linear scale in the box to
the box size;  as this ratio becomes larger so does the mass of
the largest tree.  Furthermore, the amount of computation required for a
tree with $N_t$ particles will scale as $N_t$log$N_t$. 
Thus, there can be a substantial variation in the computational
time required between different trees, and evolving the largest 
tree can comprise a significant fraction of the total computational
load.  An efficient parallel code must handle this situation well
when dividing work up among NCPU processors.

Load balancing of trees is achieved in three ways.  First, trees are sorted
by $N_t$ and then divided into ``copses'' of roughly equal
amounts of work using the ``greedy'' algorithm.  
That is, starting with the largest tree, each one in turn is
given to the copse which up to that point has been assigned the 
least amount of total work.
Usually there is one copse per CPU, but there can be two or more per CPU
if required by space constraints.  There is a communication step,
when all data associated with a copse is sent to one CPU.
A CPU then evolves each
tree in its local copse in turn, starting with the most massive.
Additionally, the largest trees can be done in parallel.  If a few
large trees dominate the work load, then it is impossible for all
copses to contain equal amounts of work--- ideally, one would want
NCPU copses, each comprising 1/NCPU of the total amount of work, but
this clearly can't happen if the largest tree takes a larger fraction
just by itself.  In this case, the
force calculation for these large trees is done in parallel
by a small number of CPUs.  This is currently done quite
crudely, with only the tree walk and force calculation actually
carried out in parallel; in theory a fully parallel tree code
could be used for every tree, allocating more processors
to those copses containing the most work.
Only a few nodes are usually required to reduce the time
spent on the largest tree to the level required for load balancing.
As a final means of balancing the load,
when a CPU finishes evolving all the trees in its
local copse, it then sends a signal to the other CPUs.  A CPU with
work still left will send an unevolved tree to the idle CPU
for it to carry out the evolution.

\begin{figure}
\plotone{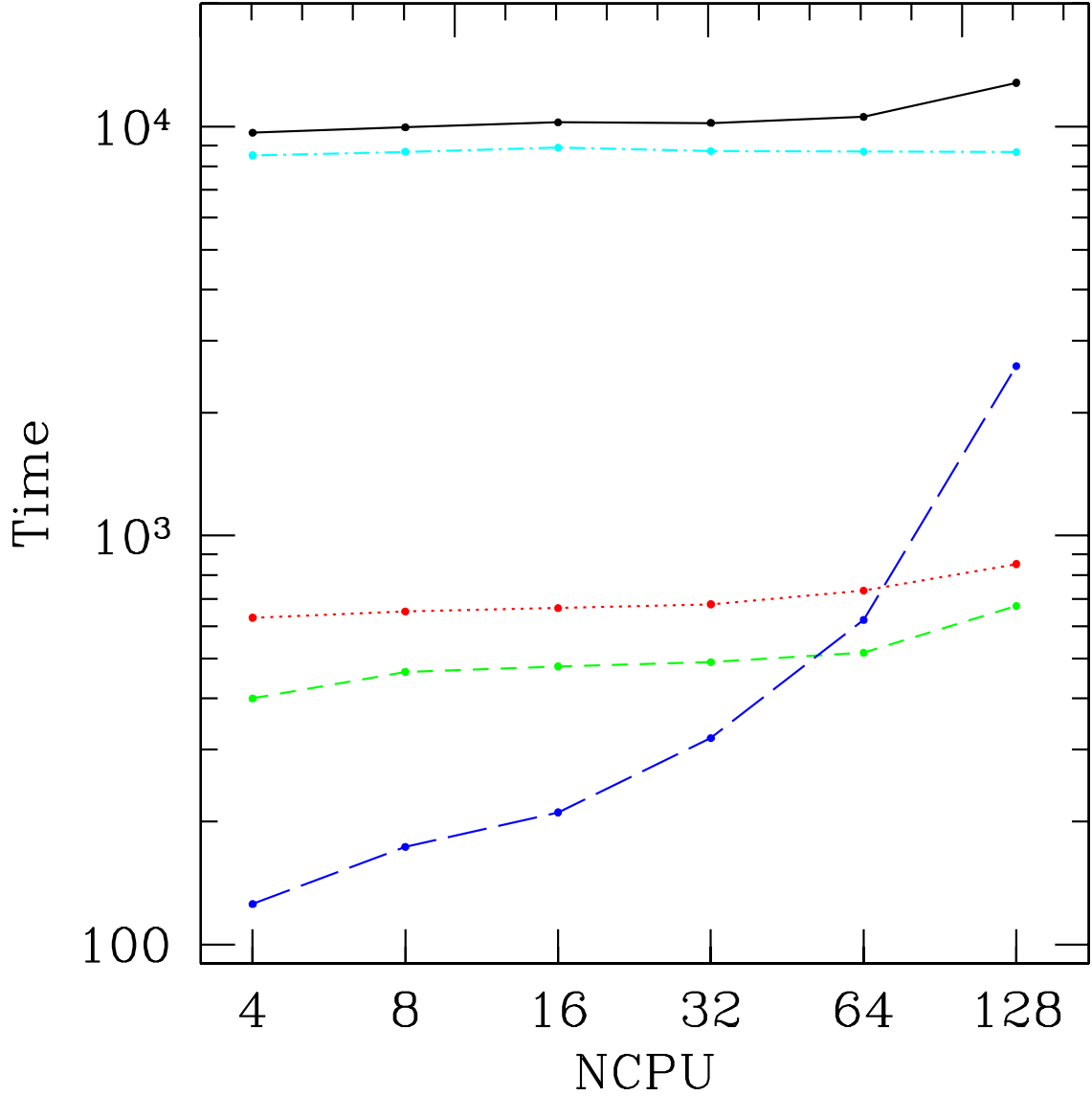}
\caption{Scaling with number of processors, NCPU. The time shown
is wallclock time in seconds multiplied by NCPU, so perfect scaling
would be a horizontal line. From top to
bottom at NCPU=4: total, tree evolution, PM, domain decomposition plus
other tree overhead, and time in communication plus load imbalance.
The test case is standard LCDM at redshift $z$=0.16, with $N$=256$^3$ particles
in a 320$h^{-1}$Mpc cube.
\label{fig:scale}}
\end{figure}

The scaling of the current implementation of TPM with 
number of processors is shown in Fig$.$\,\ref{fig:scale}.
The test case for these timing runs is a standard LCDM
model at redshift $z$=0.16 with $N$=256$^3$ particles 
in a 320$h^{-1}$Mpc cube.
This particular set of runs was carried out on an
IA-64 Linux cluster at NCSA named ``Titan''.  This machine consists
of 128 nodes, each with dual Intel 800MHz Itanium processors, and a
Myrinet network interconnect.  While the total time required depends
on processor performance, similar scaling with NCPU has been found
on a number of machines with various types of processors and
interconnects.  The topmost line in Fig$.$\,\ref{fig:scale}
is total time, calculated as the number of seconds wallclock time 
per PM step multiplied by NCPU;  perfect scaling would be a horizontal
line.  TPM performs quite well---  at NCPU=64 the efficiency is
still 91\% as compared to NCPU=4.  Beyond this point scaling begins
to degrade,  with the efficiency dropping to 75\% for NCPU=128.

The reason TPM scales well can be seen in the second line from the
top, which shows the total time spent in tree evolution.  This part
of the code takes up most of the CPU time, but it requires no
communication and there are enough trees so that just 
coarse--grained parallelization works reasonably well.  
The next two curves shown indicate the amount of
time in the PM portion of the code and overhead related to trees
(identifying tree regions and particles, etc.).  These take a small
fraction of the total time and 
scale well since the grid is distributed across all processors.
The final
curve in Fig$.$\,\ref{fig:scale} shows the time related to 
communication and imbalance in the tree part of the code.
As NCPU increases it becomes more difficult to divide the tree
work evenly and processors spend more time either recruiting work
from others or waiting for them to finish. 
This overhead could
likely be reduced by incorporating a fully parallel tree code
and also by computing the nonperiodic FFT (required to find the
tidal potential--- see \boxpap ) in parallel.
A run with a larger number of particles and grid points, and thus
a larger number of trees, would show efficient scaling beyond
NCPU=128.

\section{Performance and Accuracy} \label{sec:perfandacc}

\subsection{Spherical Overdensity Test} \label{sec:sphodtest}

As a test of new code we have
carried out a simulation of the 
secondary infall and accretion onto an initially uniform
overdensity.  This can be compared both to other codes
and to the analytic, self-similar 
solution of \citet{Bert85}.  To create
the initial conditions, $64^3$  particles were placed on
a uniform grid with zero velocity.  The eight particles
on the grid corners were then removed and placed at the
center of the volume with a spacing one half that of the
regular grid.  Thus, this is actually more of a cubic 
overdensity than spherical, but it quickly collapses and
the subsequent infall is
independent of the details of the initial state.
Also, there is a void located
half the box distance away from the overdensity,  but 
the evolution is not carried out for a long enough time
for this to be significant.  The 
initial condition is integrated from expansion factor
$a=0.01$ to $a=1$.  The values $A=2$ and $B=10$ were used
for the constants in eq.~[\ref{eqn:rhocrit}].  Since only one
halo is forming in the box, $\sigma$ remains small, rising
to only 0.5 by the end of the run.  At the end, there are
roughly 500 particles within the turnaround radius.
This test can be seen as exploring how well the initial
collapse of small objects is followed, which is the beginning
stage of halo formation in an hierarchical scenario.

\begin{figure}
\plotone{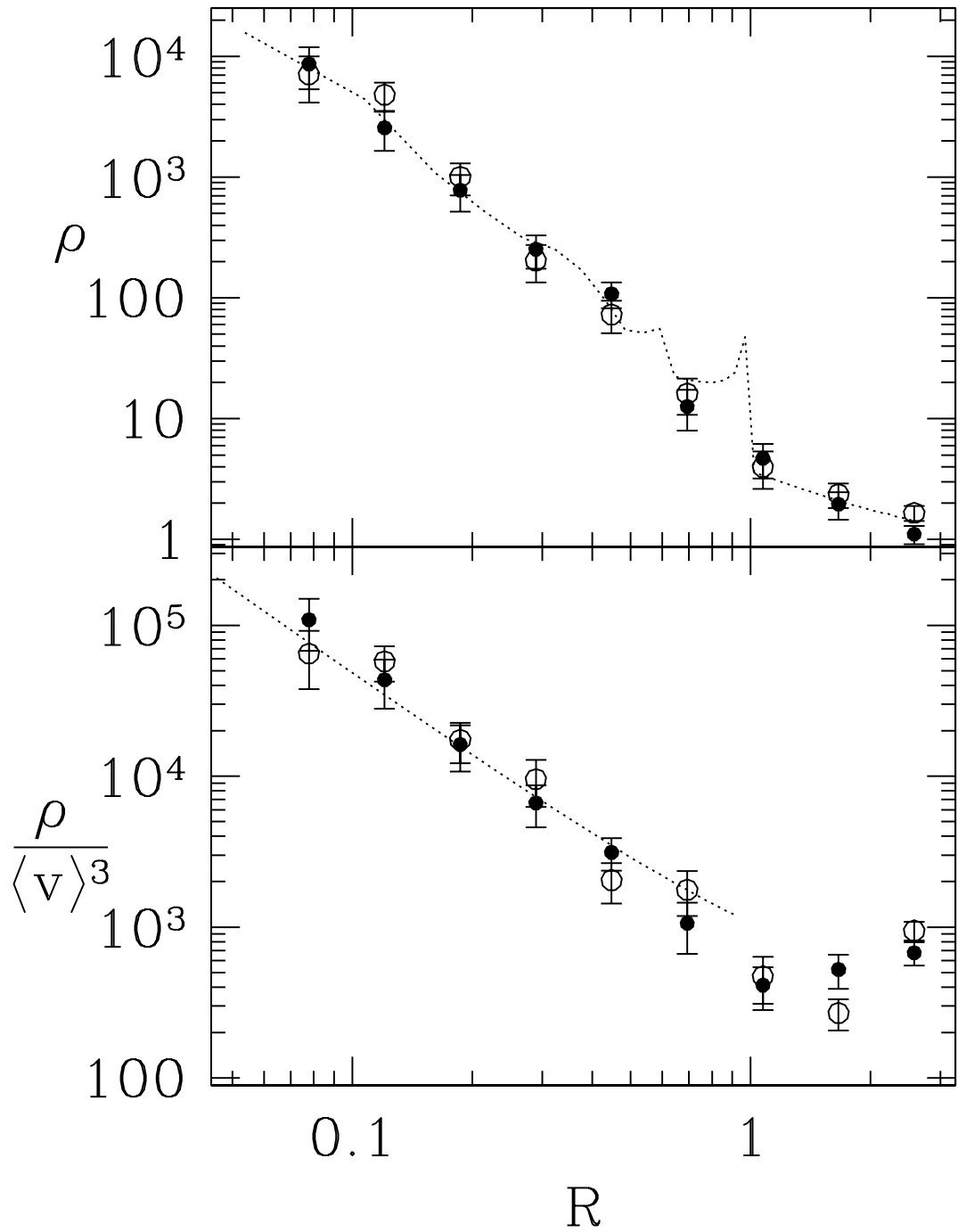}
\caption{The final state of the spherical overdensity infall test.
Filled circles: TPM run; open circles: P$^3$M run; dotted lines:
Bertschinger's analytic solution.
\label{fig:sphod}}
\end{figure}

The final state of this run is shown in Fig$.$\,\ref{fig:sphod}.
The top panel shows density as a function of radius.  Filled
points are the TPM model; error bars are the square root
of the number of particles in a given radial bin.  The inner
edge of the innermost bin is twice the softening length.
Also shown is the same run carried out with a P$^3$M code
(as open circles; details of this code are discussed
in \S\ref{sec:codecomp}), and the solution of \citet{Bert85}. 
The agreement is quite good--- the main limitation of this
run is likely the mass resolution.
As a measure of the phase space density, 
the lower panel shows $\rho/\langle v \rangle^3$, where 
$\langle v \rangle$ is the velocity dispersion of the 
particles in each radial bin.  Following \citet{TaylNava01},
we compare this to the \citet{Bert85} solution for a
$\gamma=5/3$ gas.  Again the agreement is quite reasonable.
Differences between TPM and P$^3$M arise from different types
of softening (P$^3$M uses Plummer softening; here $\epsilon$
was set to half the TPM value) and from time stepping
(for P$^3$M, $\Delta t$ for all particles was set by the
minimum $\sqrt{0.05\epsilon/g}$ ).

\subsection{Comparison with other codes} \label{sec:codecomp}

As a test of TPM in a less idealized situation, the initial
conditions described in \S\ref{sec:overv} were again evolved,
but using two other N-body codes.  The first is the P$^3$M code 
of \citet{FerrBert94} in a version made parallel by \citet{Fred97}---
this code was also used in \S\ref{sec:sphodtest}.  
As in the TPM run, a mesh of $256^3$ grid 
points was used. This code uses Plummer softening; the softening length 
was set to half the value used in the TPM run.
The time variable in this code is $\tau$, defined
by $d\tau = dt/a^2(t)$; all particles have the same time step,
set by the minimum $\sqrt{0.05\epsilon/g}$.
The other N-body code used is the tree code named
GADGET\footnote{Publically available at
\url{http://www.MPA-Garching.MPG.DE/gadget/}}
of \citet{SpringelYW01}. 
In this code,
periodic boundary conditions are handled with Ewald summation,
the time variable is the expansion factor $a$,
and each particle has an individual time step.
Following \citet{Powetal02}, the conservative tree node opening
criterion flagged by -DBMAX was used, and the time step was set
by $\sqrt{0.03\epsilon/g}$. In addition,
the maximum allowed time step was set to 1\% of the initial
expansion factor;  whenever $a$ doubled this maximum was 
also increased by a factor of two, by checkpointing the
run and restarting with the new value.
The softening length was set to be the same as the TPM
run (allowing for different notational conventions).

\begin{figure}
\plotone{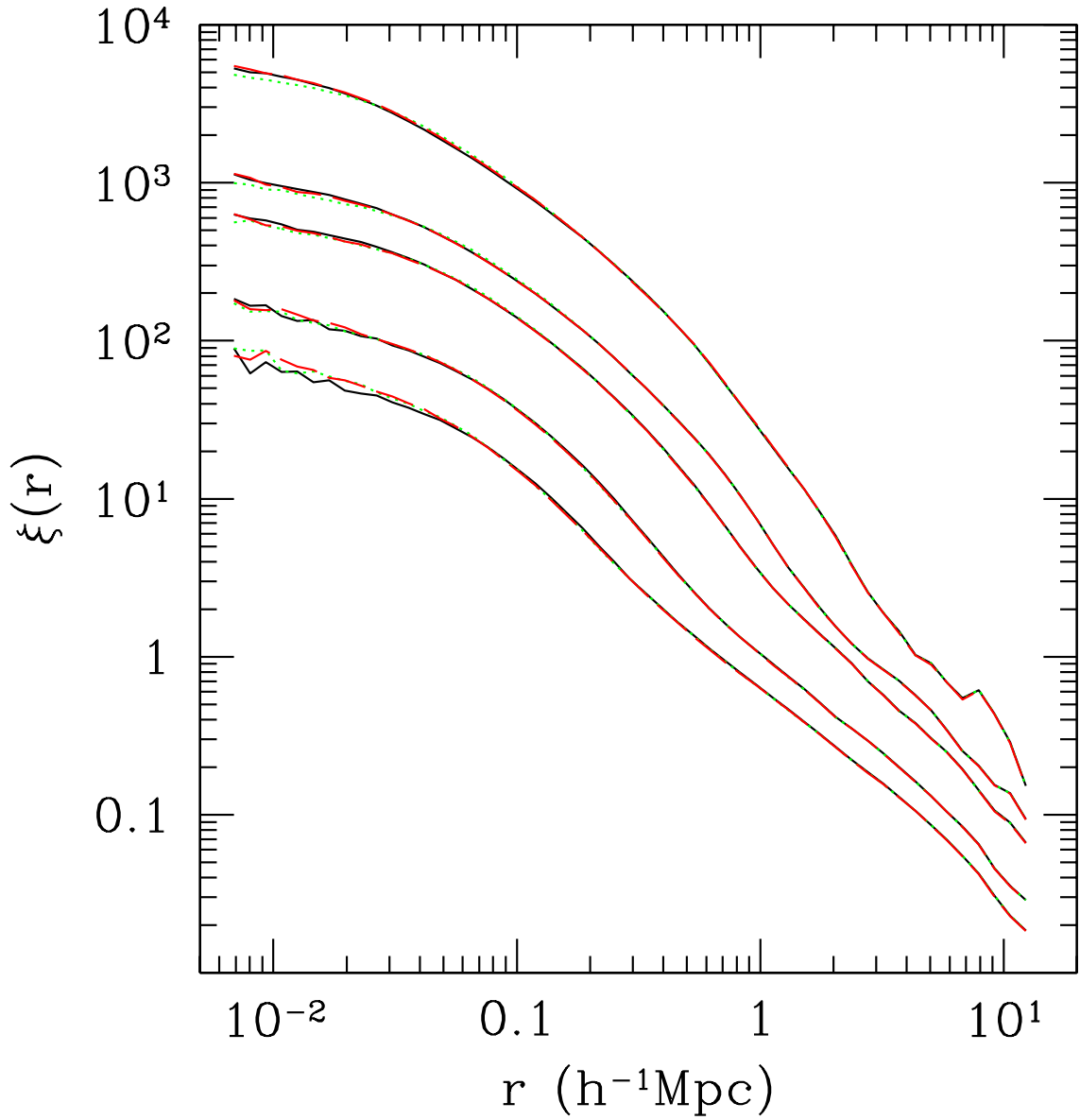}
\caption{Particle--particle correlation function for TPM (solid), 
P$^3$M (dotted), and tree (dashed lines) codes. From top to bottom: 
$z$ = 0, 1, 2, 3, and 4. 
\label{fig:cfcmp}}
\end{figure}

As a first comparison of the codes, the two point correlation 
function $\xi(r)$, found by counting the number of particle pairs in 
bins of separation $r$, was calculated. 
To compute $\xi(r)$, 51 logarithmically spaced bins 
were used, with the minimum pair separation considered being
$2\epsilon=6.4h^{-1}$kpc and the maximum one third of the
box size.  The results at various redshifts are shown
for all three runs in Fig$.$\,\ref{fig:cfcmp}.
Clearly there is little difference to be seen in the three
codes.  At larger scales this is to be expected; at smaller
scales force approximations, smoothing, and relaxation may become important.
The fact that P$^3$M uses Plummer rather than spline softening
explains why it shows a lower $\xi(r)$ at scales less than
a few times the softening length.
If high values of $\xi(r)$ provide a measure of accuracy, the
TPM and GADGET are slightly more accurate than P$^3$M.

\begin{figure}
\plotone{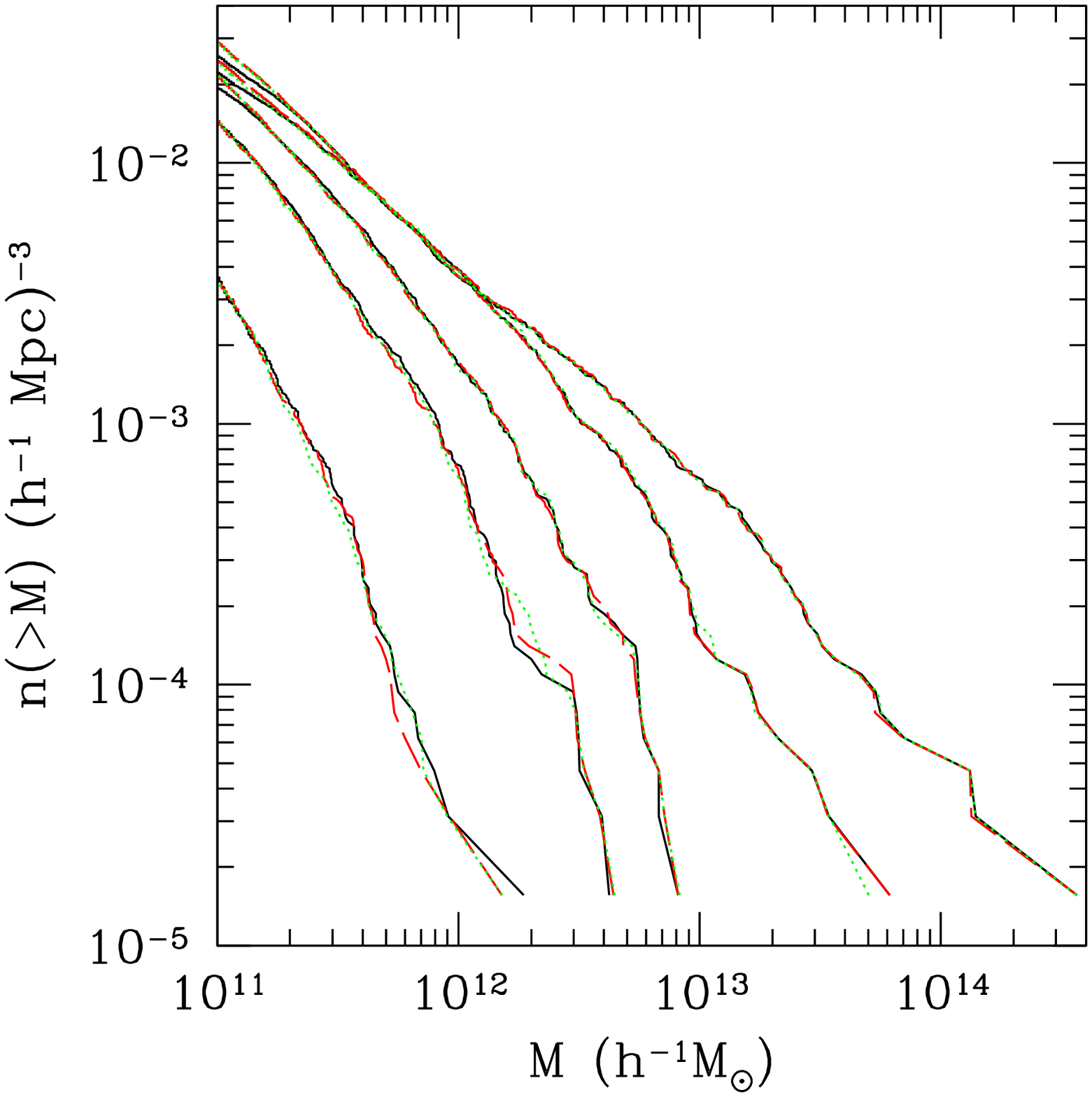}
\caption{Cumulative FOF halo mass function for TPM (solid), P$^3$M (dotted),
and tree (dashed) codes. From top to bottom: $z$ = 0, 2, 3, 4, and 6. 
\label{fig:fofcmp}}
\end{figure}

The statistics of mass peaks--- the dark halos surrounding galaxies,
clusters, and so on---  are an important product of N-body codes
\citep{White02}.
One commonly used halo finder is friends--of--friends, or 
FOF\footnote{We employed the University of Washington NASA HPCC ESS 
group's FOF code, available at 
\url{http://www-hpcc.astro.washington.edu/}} \citep{DavisEFW85}.
The cumulative mass function, found with 
FOF using a linking length of 0.2 times the mean interparticle
separation, is shown in Fig$.$\,\ref{fig:fofcmp}.  At higher redshifts
there is little discernible difference between the three
codes.  At later times, TPM
seems to have fewer halos containing $\lesssim$50 particles.
It is unclear how to interpret this, because halos with less
than 150 particles (i.e. mass $<3.8\times 10^{11}h^{-1}M_\odot$)
are affected by two-body relaxation.  
\citet{KnebeGB01} found that an adaptive P$^3$M and GADGET both
produced more small FOF halos than the adaptive multigrid
MLAPM and ART codes, so in this case TPM may agree more closely
with the latter.

\begin{figure}
\plotone{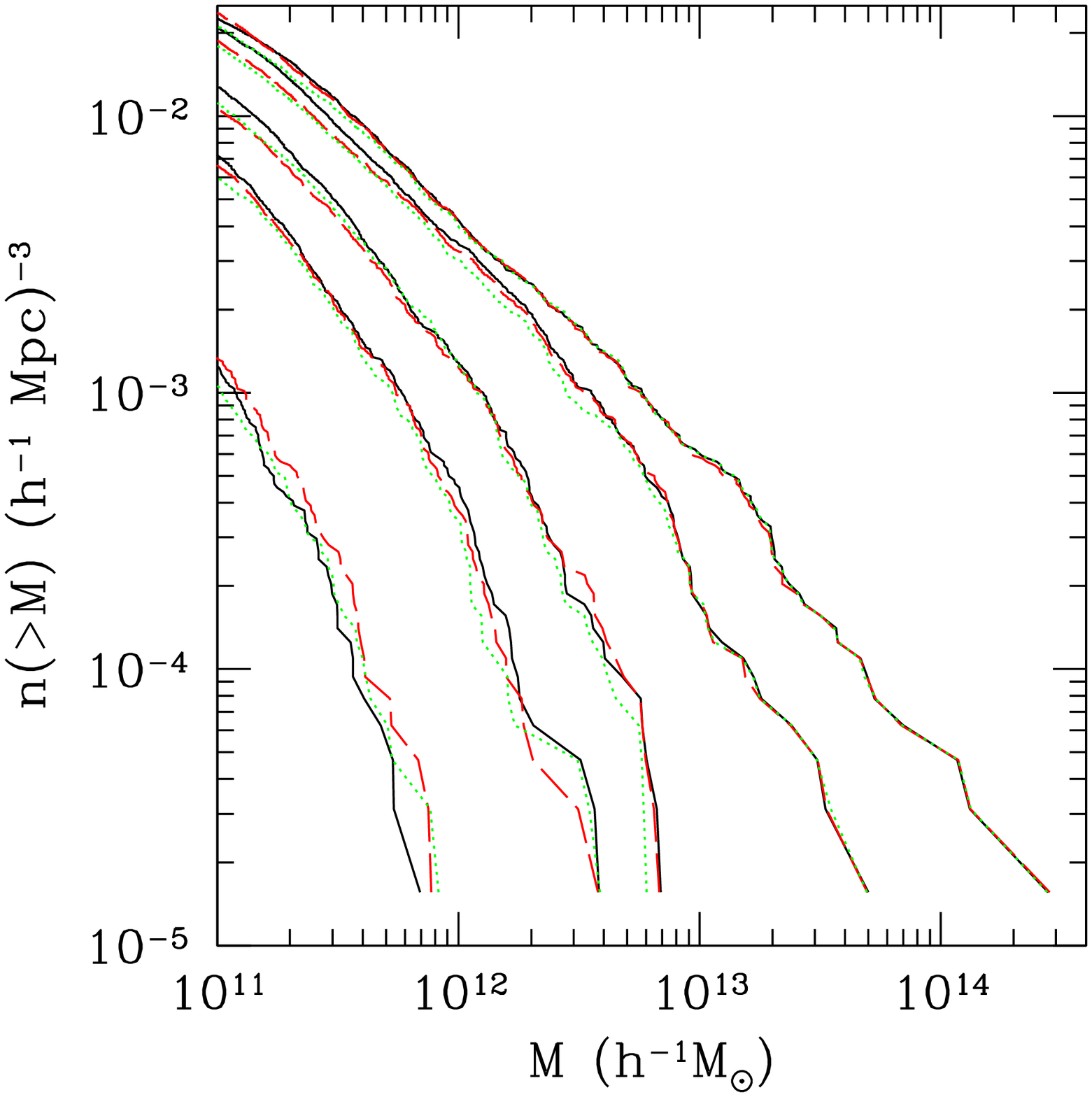}
\caption{Cumulative BDM halo mass function for TPM (solid), P$^3$M
(dotted), and tree (dashed) codes.
From top to bottom: $z$ = 0, 2, 3, 4, and 6. 
\label{fig:mfcmp}}
\end{figure}

To investigate further a different halo finder was used, namely
the Bound Density Maxima (BDM) algorithm\footnote{Publically available 
at \url{http://astro.nmsu.edu/$\sim$aklypin/}} \citep{KlypGKK99}.
This algorithm is significantly different
from FOF, so the resulting mass function is probing different
qualities of the dark matter distribution.
Density maxima inside spheres of radius 100 $h^{-1}$kpc
were found, and halo centers were then calculated using 
spheres of radius 40 $h^{-1}$kpc.  The mass for each halo is
taken to be that inside a sphere containing the overdensity
expected for a virialized halo just collapsed from a spherical
top--hat;  to calculate this overdensity the fit of \citet{BryNor98}
was used. In BDM (unlike FOF) particles moving faster than
the escape velocity are removed from the halo.  
The BDM cumulative mass functions
for the three N-body codes are shown in Fig$.$\,\ref{fig:mfcmp}.
As with FOF, the agreement between codes is quite good.
In this case it appears to be TPM which produces more small
halos, but again this effect is for low mass halos likely affected by
relaxation.

\begin{figure}
\plotone{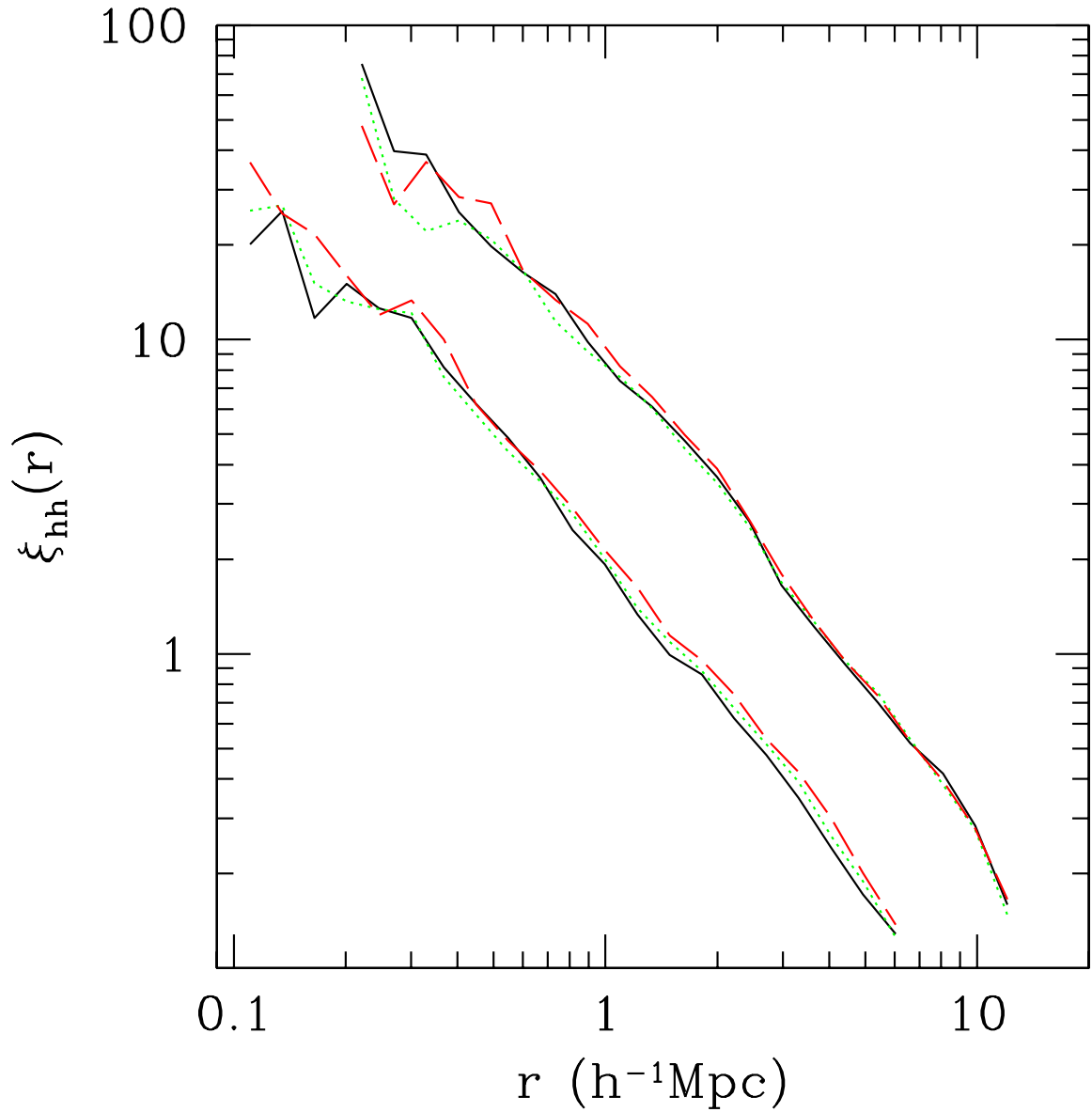}
\caption{The halo--halo correlation function for all halos with
40 particles or more (mass $>10^{11}h^{-1}M_\odot$)
at $z$=0 and 1 (using physical radius);
TPM (solid), P$^3$M (dotted), and tree (dashed) codes.
\label{fig:cfhal}}
\end{figure}

Using the BDM halos with more than 40 particles, the halo--halo
correlation function is shown in Fig$.$\,\ref{fig:cfhal}.
The number of pairs of halos in 21 logarithmically spaced radial bins,
with the smallest separation being 200 $h^{-1}$kpc and the largest
one third of the box size, were tallied and compared with the
expectation for a random distribution.  All three codes give
the same result.  From the results presented so far in this section
it is clear that TPM yields a quite similar evolved matter distribution
as compared to other codes.

\begin{figure}
\plotone{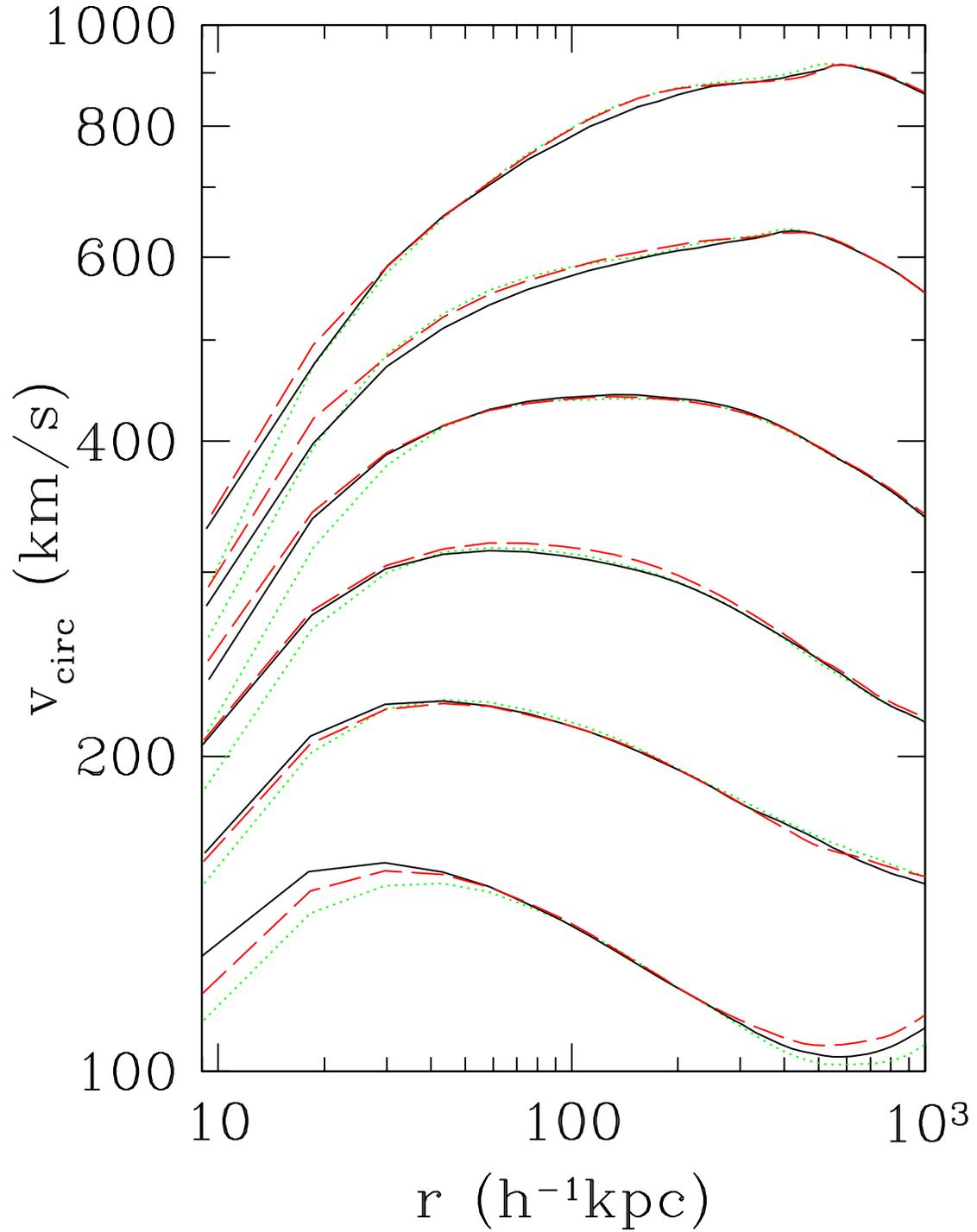}
\caption{Average circular velocity profiles for TPM (solid), P$^3$M (dotted),
and tree (dashed) codes.  Each bin spans a factor of $\sqrt{10}$, and 
the lowest mass used is $3.81\times 10^{11}h^{-1}M_\odot$ (150 particles).
\label{fig:vcirc}}
\end{figure}

Finally we turn to the internal properties of halos.
One probe of the mass distribution in a halo is the
circular velocity $v_c\equiv \sqrt{GM(<r)/r}$.  For each simulation we
divide halos up into six mass bins.  To avoid relaxation
effects, the lowest mass considered is $3.81\times 10^{11}h^{-1}M_\odot$.
The width of each bin is a factor of $\sqrt{10}$, so the largest
mass bin contains two halos with mass above 
$1.20\times 10^{14}h^{-1}M_\odot$.  For each set of halos the average
circular velocity as a function of radius was calculated;  these 
average velocity curves are shown in Fig$.$\,\ref{fig:vcirc}.
It can be seen that
these curves are quite similar for all three codes.
Two main trends are noticeable, the main one being that P$^3$M
shows lower circular velocities at small radii.
This is likely due to the use of Plummer softening and less
strict time step criterion in the P$^3$M run, both of which would lead
to lower densities in the innermost parts of halos. 
The other main difference is that
for the lowest masses it appears that TPM yields higher
circular velocities than the other two codes. 

Instead of averaging over a number of halos of similar mass, it
is also possible to compare halos on a one-by-one basis.
To find the appropriate pairs of halos, the list of halos 
with more than 150 particles found by BDM was sorted by mass, 
and for each target GADGET or P$^3$M halo 
a TPM halo was selected as a match if its mass was within 25\%
and its position within 1 $h^{-1}$Mpc of the target halo; if more
than one halo passed this test then the nearest in
position was selected.
The selected TPM halo was removed from further consideration, and
the process repeated for next target halo.
In this manner a match was found for 588 out of 609 GADGET halos
and 573 out of 582 P$^3$M halos.  For various measured halo 
properties, the percentage difference of the TPM halo from the 
GADGET or P$^3$M halo was calculated.  Since the dispersion of
these differences increases as less massive halos are considered,
halo pairs are split into two groups, with the higher mass
group containing all target halos with 394 or more particles
(i.e. mass $\geq 10^{12}h^{-1}M_\odot$).

Results are shown in Table ~\ref{tab:hdiff}, which for each
property gives the
mean percentage difference and one standard deviation, as well
as the first quartile, median percentage difference, and third
quartile.  The first property shown is halo mass $M$.  The
difference here is constrained to be less than 25\%, but in 
most cases the TPM value is within 10\% of the other code's.
It appears that P$^3$M yields slightly lower masses than the
other two codes, but the latter two agree quite well.  The agreement
in masses shows that the simple scheme used to find matching
halo pairs works well, as does the fact that the difference
in position is less than 165$h^{-1}$kpc in 95\% of the cases.
This is reinforced by the agreement between the halo center of
mass velocities $v_{cm}$ shown in Table ~\ref{tab:hdiff}; the
velocity vectors are closely aligned, differing by less than
$7.5\degr$ in 95\% of the pairs.  Thus only a few percent of the 
halo pairs are mismatches; since these pairs still have
similar masses and are likely in similar environments, they
are kept in the comparisons.

In terms of the 3-D root mean square velocity $v_{rms}$ and maximum 
circular velocity $v_c$,
the more massive halos are very similar in all three codes,
with no offset between codes.  However, at the lower
mass end TPM gives values that tend to be systematically
higher by a few percent.  This can also be seen for the average 
circular velocity in the bottom curve of Fig$.$\,\ref{fig:vcirc};
this is curve is the average for halos below 
roughly $10^{12}h^{-1}M_\odot$.

The difference between TPM and the other two codes is most
pronounced near halo centers.  The last property shown in 
Table ~\ref{tab:hdiff} is a comparison of central density $\rho_o$,
which is computed by measuring the amount of mass within
$4\epsilon$=12.8$h^{-1}$kpc of the halo center.  As would be expected 
from Fig$.$\,\ref{fig:vcirc}, TPM yields higher central densities
than GADGET, with P$^3$M giving lower $\rho_o$ than either of
these.

\begin{figure}
\plotone{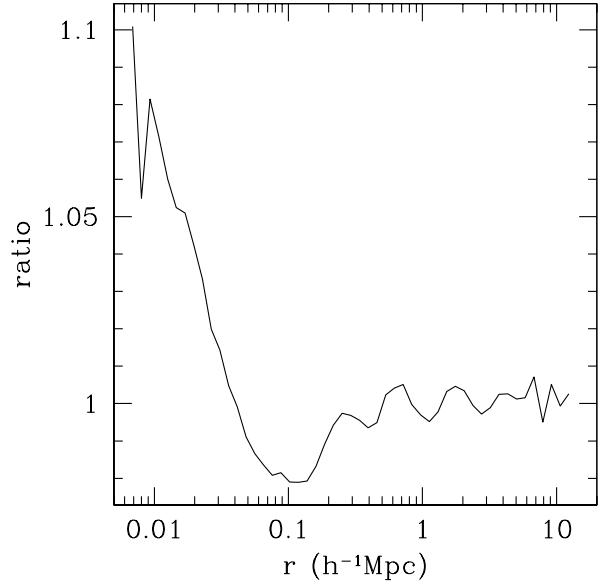}
\caption{Ratio of the particle--particle correlation function
$\xi(r)$ for the standard TPM run (using eq.~[\ref{eqn:deltat}]) 
to that for the run with a less stringent time step criterion.
\label{fig:xicmp}}
\end{figure}

One possible source of the differences seen in TPM halos
as compared to a pure tree code could simply be the choice of
time step.  The P$^3$M run took $2.1\times10^4$ steps and the
GADGET run $2.8\times10^4$; TPM on the other hand took 
$4.4\times10^4$, or 60\% more than GADGET.  Of course this
comparison is not entirely straightforward because different
particles determine the smallest time step at different times.
Part of the difference between GADGET and TPM may be due to
the fact that TPM bins time steps by factors of two, whereas
GADGET allows time steps to vary more gradually.
A second TPM run was carried out to separate other code differences 
from the time step criterion. This
run was identical to the first except that the time step was
set not by eq.~[\ref{eqn:deltat}], but rather  
by $\sqrt{0.05a^3\epsilon/g}$ (similar to the GADGET and P$^3$M codes); 
as a result it took fewer
steps than any of the other runs.  There is no significant
difference in the mass functions and halo--halo correlation
functions between this run and the original TPM run. 
The particle-particle correlation function is different, however. 
This can be seen in Fig$.$\,\ref{fig:xicmp}, which shows the 
ratio of the original TPM run's $\xi(r)$ to that of the new run,
at redshift $z$=0.
The new TPM run has a lower $\xi(r)$ for $r<20h^{-1}$kpc, roughly
5-10\% lower than the original TPM and similar to the P$^3$M run.
This indicates the internal structure of halos has
been affected by the longer time steps, which is confirmed by
repeating the comparison of individual halos.  The difference 
between halos in the new TPM run and the GADGET run are given in
Table ~\ref{tab:hdiff}.  For the higher mass halos the new
run, unlike the original one, tends to give lower maximum circular 
velocities and central densities than the tree code.  This indicates
that the longer time step is in fact leading to inaccuracies.
The differences between TPM and the tree code for low mass halos
seen in the original run
persist in the new run, though they are not quite as pronounced.

\begin{figure}
\plotone{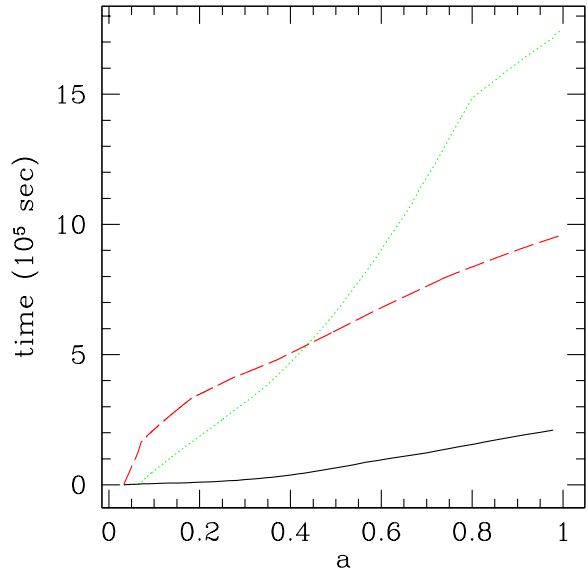}
\caption{Wall-clock time (in units of $10^5$ seconds) used by
TPM (solid), P$^3$M (dotted), and tree (dashed) codes as a function
of expansion parameter for a standard LCDM model.  All codes were
run on four processors of an SGI Origin 2000.
\label{fig:times}}
\end{figure}

Another point of comparison between codes is efficiency with which
they use computing resources.  Fig$.$\,\ref{fig:times} shows the
wall-clock time consumed by each of the codes,
as a function of expansion parameter,
in carrying out the
test simulation.  All the runs used four 300 MHZ IP27 processors 
of an SGI Origin 2000; the codes were compiled with the SGI compilers
and MPI library.  All three codes used roughly the same amount
of memory (TPM requires at least 20 reals per particle plus 3 reals
per mesh point, divided evenly among processors).  
At the earliest times both P$^3$M and TPM spend
most of their time in the PM FFT, and so they behave similarly.
However, as objects begin to collapse P$^3$M begins to consume more time
computing particle-particle interactions.  The fact that the
accelerations of all particles are updated every step also makes
P$^3$M use more time than do the two multiple time step codes.
The tree code takes more time when the particle distribution is
nearly homogeneous, demonstrating that a PM code (which is what
the gridded codes basically are in this situation) is very
efficient.  However, the tree code timing is roughly independent
of the particle distribution, and once inhomogeneity develops
it does not require more time per update, whereas the other codes
do.  TPM does well compared to the tree code for a couple 
of reasons.  By $a$=1 roughly half the particles in the TPM run
are still being handled solely by PM.  Also, imagine breaking $N$
particles up into $t$ trees each with $N/t$ particles;  the time
required to solve all the trees will scale as $N\log{N/t}$,
lower by a logarithmic factor than using the same tree solver
on all the particles.
Overall, despite taking more timesteps, in this test case TPM required
less CPU time than the other codes, by a factor of $\sim 4$ relative
to the tree code and $\sim 8$ relative to P$^3$M.

\section{Conclusions} \label{sec:discussion}

This paper has presented a parallel implementation of the TPM algorithm. 
Several improvements over the
implementation of \boxpap\  have been made.  Particles in tree 
regions each have an individual time step, half of
the PM time step or less, making the tree integration more efficient. 
The treatment of tidal forces on trees is also improved by saving
the tidal potential on a grid and evaluating the force as a
function of particle position at each smaller particle time step;
thus a greater ratio of tree to PM time steps is allowed.
A new, more stringent, time step criterion has been implemented.
A new criterion for locating regions for treatment with higher
resolution is given; by finding cells with higher density than
their surroundings, small halos in lower density regions are located
and followed at full resolution.
These changes significantly increase the speed and accuracy of TPM:
a computation of the test case discussed in \S\ref{sec:codecomp} using 
the code described in \boxpap\ required more CPU time (by a 
factor of over three) than the current version, but was
only accurate for halos with more than 315 particles.

Comparisons with other widely used algorithms were made for a 
typical cosmological structure formation simulation.  These show
excellent agreement.  The particle-particle correlation functions,
the halo mass functions, and the halo-halo correlation functions
from TPM agree quite well with those from P$^3$M and tree codes.
The internal properties of halos also agree; the main difference
being that, for lower mass halos, TPM yields higher $rms$ and
maximum circular velocities (by a few percent over a tree code). 
TPM halos also show higher central densities than those of the other
two codes, though the mean difference is smaller than the dispersion.
This difference disappeared, at least among the more massive halos,
when the time step criterion of eq.~[\ref{eqn:deltat}] was replaced
with one similar to that employed by the other two codes.
Thus we conclude that a choice of a relatively conservative time
step criterion contributed to a slightly improved accuracy.

TPM yielded results of similar accuracy to the other codes 
used here while using significantly less computational time
(about a quarter of that needed by a tree code and an eighth
of that needed by P$^3$M).
It also scales well
on distributed memory parallel machines, such as networked PCs,
because this parallelism is built in as part of the design of TPM.
However, in committing to an algorithm which accentuates the 
coarse-grained parallelism inherent in a typical cosmological
simulation, a large degree of flexibility is sacrificed.  A basic
presumption of TPM is that the largest nonlinear structure inside
the simulation box is a small fraction of the total mass and
volume.  To simulate a situation where this is not the case
(e.g. two colliding galaxies) another code would be preferred.

The TPM source code can be obtained at 
\url{http://astro.princeton.edu/$\sim$bode/TPM} 
or by contacting the authors.
The code is written in Fortran 77 and uses MPI for message passing;
thus it is very portable and can be used on clustered PC's or
other distributed memory systems.

\acknowledgments

Many thanks are due to Lars Hernquist for generously supplying
a copy of his tree code; also Edmund Bertschinger for use of his
P$^3$M code, Joe Henawi for help with FOF, and Scott Tremaine
for useful discussions.
This research was supported by the National Computational Science Alliance 
under NSF Cooperative Agreement ASC97-40300, PACI Subaward 766.
Computer time was provided by NCSA and the Pittsburgh Supercomputing
Center.

%%%\begin{thebibliography}{}

% try:

%%%\clearpage

\begin{table*}
\begin{center}
\caption{Halo comparison --- percentage differences
\label{tab:hdiff} }
\begin{tabular}{lccccccccccc}
\tableline\tableline
& \multicolumn{5}{c}{$3.8 \times 10^{11} < M < 10^{12}$ } & & 
  \multicolumn{5}{c}{$M \geq 10^{12}  \  h^{-1}M_\odot$ } \\
\cline{2-6}
\cline{8-12}
 & mean & s.d. & $Q_1$ & $Q_2$ & $Q_3$ & 
 & mean & s.d. & $Q_1$ & $Q_2$ & $Q_3$   \\
\cline{2-6}
\cline{8-12}
\multicolumn{12}{l}{TPM vs. Tree} \\
$M$       & -0.03 & 8.21 & ~-5.07 & 0.08 & 4.80 & & -0.37 & 5.33 & ~-2.33 & -0.12 & 2.08 \\
$v_{cm}$  & 0.06  & 9.38 & ~-1.08 & 0.52 & 1.74 & &  0.49 & 4.28 & ~-0.54 &  0.59 & 1.74 \\
$v_{rms}$ & 1.59  & 7.94 & ~-2.20 & 1.99 & 5.97 & & -0.05 & 4.38 & ~-2.12 &  0.18 & 2.23 \\
$v_c$     & 2.69  & 6.98 & ~-1.14 & 2.89 & 6.68 & &  0.35 & 3.85 & ~-1.48 &  0.25 & 2.01 \\
$\rho_o$  & 16.2  & 38.1 & ~-2.22 & 8.33 & 26.0 & &  4.03 & 22.1 & ~-8.82 &  2.45 & 13.7 \\
\multicolumn{12}{l}{TPM vs. P$^3$M}\\
$M$       & 1.84 & 8.65 & ~-3.40 &  1.75 & 6.74 & &  0.38 & 5.14 & ~-1.51 &  0.39 & 2.37 \\
$v_{cm}$  & 0.22 & 7.44 & ~-1.19 & -0.10 & 1.20 & & -0.38 & 4.26 & ~-0.93 & -0.07 & 0.84 \\
$v_{rms}$ & 3.84 & 7.13 & ~-0.80 &  3.59 & 8.94 & &  0.23 & 3.82 & ~-1.83 & -0.03 & 2.16 \\
$v_c$     & 6.20 & 8.93 &  ~1.02 &  5.11 & 10.7 & &  0.32 & 4.06 & ~-1.94 & -0.24 & 2.00 \\
$\rho_o$  & 29.4 & 39.2 &  ~3.45 &  20.0 & 47.2 & &  13.7 & 35.0 & ~-2.46 &  6.82 & 19.0 \\
\multicolumn{12}{l}{TPM ($\Delta t<\sqrt{0.05a^3\epsilon/g}$) vs. Tree}\\
$M$ & -0.25 & 8.11 & ~-4.70 & -0.08 & 4.26 & & -0.63 & 5.82 & ~-2.73 & -0.33 & 2.18 \\
$v_{cm}$ & -0.01 & 8.19 & ~-0.97 & 0.34 & 1.65 & & 0.76 & 7.10 & ~-0.68 & 0.45 & 1.76 \\
$v_{rms}$ & 1.55 & 7.83 & ~-3.07 & 1.99 & 5.36 & & -1.00 & 3.86 & ~-2.65 & -0.62 & 1.28 \\
$v_c$ & 1.74 & 6.80 & ~-2.61 & 1.50 & 5.69 & & -0.77 & 3.85 & ~-2.44 & -0.49 & 1.03 \\
$\rho_o$ & 12.7 & 30.5 & ~-6.24 & 6.89 & 26.7 & & -0.75 & 24.3 & ~-12.9 & -4.16 & 6.49 \\
\tableline
\end{tabular}
\end{center}
\end{table*}

%%\clearpage

\epsscale{0.95}
\epsscale{1.0}

\epsscale{0.9}
\epsscale{1}

\epsscale{0.9}
\epsscale{1}

\epsscale{0.9}
\epsscale{1}

\epsscale{0.5}

\begin{thebibliography}{}\setlength{\itemsep}{-2mm}
\bibitem[Aarseth(1999)]{Aars99} Aarseth, S. 1999, \pasp, 111, 1333
\bibitem[Baertschiger, Joyce \& Labini(2002)]{BaeJoLa02} Baertschiger, T.,
      Joyce, M. \& Labini, F.S. 2002, \apjl, 581, 63 (astro-ph/0203087)
\bibitem[Bagla(1999)]{Bagla99} Bagla, J.S. 1999, preprint (astro-ph/9911025)
\bibitem[Barnes(1998)]{Barnes98} Barnes, J.E. 1998, Galaxies: Interactions 
      and Induced Star formation, R.C. Kennicutt Jr., F. Schweizer 
      and J.E. Barnes, Berlin: Springer, 275
\bibitem[Barnes \& Hut(1986)]{BarnHut86}Barnes, J. \& Hut, P. 1986, 
      Nature, 324, 446
\bibitem[Becciani \& Antonuccio-Delogu(2001)]{BAD01}  Becciani, 
      U. \& Antonuccio-Delogu, V. 2001, Comp. Phys. Comm., 136, 54
\bibitem[Bertschinger(1985)]{Bert85} Bertschinger, E. 1985, \apjs, 58, 39
\bibitem[Bertschinger(1998)]{Bert98} Bertschinger, E. 1998, \araa, 36, 599
\bibitem[Bertschinger(2001)]{Bert01} Bertschinger, E. 2001, \apjs, 137, 1
\bibitem[Bryan \& Norman(1998)]{BryNor98} Bryan, G.L. \& Norman,
      M.L. 1998, \apj, 495, 80
\bibitem[Bode, Ostriker, \& Xu(2000)]{BOX00} Bode, P., Ostriker, J.P.,
      \& Xu, G. 2000, \apjs, 128, 561 (\boxpap )
\bibitem[Calder et al.(2002)]{Calder02} Calder, A.C. et al. 2002,
       \apjs, in press
\bibitem[Couchman, Thomas \& Pearce(1995)]{CouchTP95} Couchman, H.M.P.,
       Thomas, P.A. \& Pearce, F.R. 1995, \apj, 452, 797
\bibitem[Davis et al.(1985)]{DavisEFW85} Davis, M., Efstathiou G.,
      Frenk, C. \&  White, S.D.M. 1985, \apj, 292, 371
\bibitem[Dehnen(2000)]{Dehnen00} Dehnen, W. 2000, \apjl, 536, L39
\bibitem[Dehnen(2002)]{Dehnen02} Dehnen, W. 2002, J. Comp. Phys., 179, 27
\bibitem[Dorband, Hemsendorf \& Merritt(2002)]{DoHeMe02} Dorband, E.N., 
      Hemsendorf, M. \& Merritt, D. 2002, J. Comp. Phys., 
      in press (astro-ph/0112092)
\bibitem[Efstathiou et al.(1985)]{Efst85}Efstathiou G., Davis, M.,
      Frenk, C. \& White, S. 1985, \apjs, 57, 241
\bibitem[Ferrell \& Bertschinger(1994)]{FerrBert94}Ferrell, R. \&
    Bertschinger, E.  1994, Int. J. Mod. Phys. C, 5, 933
\bibitem[Frederic(1997)]{Fred97}Frederic, J.J. 1997, Ph.D. Thesis, MIT
\bibitem[Hamana, Yoshida \& Suto(2002)]{HaYoSu02} Hamana, T.,
   Yoshida, N. \& Suto, Y. 2002, \apj, 568, 455
\bibitem[Hernquist(1987)]{Hern87} Hernquist, L. 1987, \apjs, 64, 715
\bibitem[Hernquist(1990)]{Hern90} Hernquist, L. 1990, J. Comp. Phys., 87, 137
\bibitem[Hernquist \& Katz(1989)]{HernKatz89} Hernquist, L. \& Katz, N.
    1989, \apjs, 70, 419
\bibitem[Hockney \& Eastwood(1981)]{HockEast81}  Hockney, R.W. \& Eastwood, 
    J.W. 1981, Computer Simulation Using Particles, New York: McGraw Hill
\bibitem[Kayo, Taruya \& Suto(2001)]{KayoTaSu01} Kayo, I., Taruya, 
    A. \& Suto, Y. 2001, \apj, 561, 22
\bibitem[Jing \& Suto(2002)]{JingSuto02} Jing, Y.P. \& Suto, Y. 2002, 
    \apj, 574, in press
\bibitem[Klypin(2000)]{Klypin00} Klypin, A. 2000, preprint (astro-ph/0005502)
\bibitem[Klypin et al.(1999)]{KlypGKK99} Klypin, A., Gottl\"ober, S.,
   Kravtsov, A.V. \& Khokhlov, A.M. 1999, \apj, 516, 530
\bibitem[Klypin \& Holtzman(1997)]{KlypHol97} Klypin, A., \& Holtzman,
   J. 1997, preprint (astro-ph/9712217)
\bibitem[Knebe(2002)]{Knebe02} Knebe, A. 2002, \mnras, submitted
   (astro-ph/0201490)
\bibitem[Knebe, Green \& Binney(2001)]{KnebeGB01} Knebe, A., Green, A. 
   \& Binney, J. 2001, \mnras, 325, 845
\bibitem[Knebe et al.(2000)]{KnebeKGK00} Knebe, A., Kravtsov, A.V., 
   Gottl\"ober, S. \& Klypin, A.A. 2000, \mnras, 317, 630
\bibitem[Kravtsov, Klypin \& Khokhlov(1997)]{KKK97} Kravtsov, A.V., 
   Klypin, A.A. \& Khokhlov A.M. 1997, \apjs, 111, 73
\bibitem[Lia \& Carraro(2001)]{LiaCar01} Lia, C. \& Carraro, G. 2001,
     \apss, 276, 1049
\bibitem[Ma \& Bertschinger(1995)]{MaBert95} Ma, C.-P., \& Bertschinger,
      E. 1995, \apj, 455, 7
\bibitem[Miocchi \& Capuzzo-Dolcetta(2002)]{MioCD02} Miocchi, P. \& 
     Capuzzo-Dolcetta, R. 2002, \aap, 382, 758
\bibitem[Mo \& White(2002)]{MoWhite02} Mo, H.J. \& White, S.D.M.
   2002, \mnras, 336, 112 (astro-ph/0202393)
\bibitem[Ostriker \& Steinhardt(1995)]{OstStein95} Ostriker, 
   J.P. \& Steinhardt, P.J. 1995, Nature, 377, 600
\bibitem[Power et al.(2002)]{Powetal02} Power, C., Navarro, J.F., 
   Jenkins, A., Frenk, C.S., White, S.D.M., Springel, V. Stadel, J.
   \& Quinn, T. 2002, \mnras, submitted (astro-ph/0201544)
\bibitem[Ricker, Dodelson \& Lamb(2000)]{RiDoLa00} Ricker, P.M.,
    Dodelson, S. \& Lamb, D.Q. 2000, \apj, 536, 122
\bibitem[Springel, Yoshida \& White(2001)]{SpringelYW01} Springel, V., 
     Yoshida, N. \& White, S.D.M. 2001, New Astronomy, 6, 79
\bibitem[Stadel(2002)]{Stadel02} Stadel, J.G. 2002, Ph.D. Thesis,
     University of Washington, Seattle
\bibitem[Taylor \& Navarro(2001)]{TaylNava01} Taylor, J.E. \& 
     Navarro, J.F. 2001, \apj, 563, 483
\bibitem[Teuben(1995)]{Teuben95} Teuben, P. 1995, Astronomical Data
    Analysis Software and Systems IV, R.A. Shaw, H.E. Payne \& J.J.E.
    Hayes, San Francisco: Astronomical Society of the Pacific, 398
\bibitem[van Kampen(2000)]{vanK00} van Kampen, E. 2000, \mnras,
     submitted (astro-ph/0002027)
\bibitem[Viturro \& Carpintero(2000)]{VitCar00} Viturro, H.R. \& 
     Carpintero, D.D. 2000, A\&AS, 142, 157
\bibitem[Xu(1995)]{Xu95} Xu, G. 1995, \apjs, 98, 355
\bibitem[White(2002)]{White02} White, M. 2002, \apjs, in press
\bibitem[Yahagi, Mori \& Yoshii(1999)]{YaMoYo99} Yahagi, H.,
     Mori, M. \& Yoshii, Y. 1999, \apjs, 124, 1
\bibitem[Yahagi \& Yoshii(2001)]{YaYo01} Yahagi, H. \& Yoshii,
    Y. 2001, \apj, 558, 463


\end{thebibliography}
\end{document}